\documentstyle[prb,aps]{revtex}
\gdef\tim#1{$10^{-#1}$}
\gdef\ttim#1{\,\times\,10^{-#1}\;}
\newcommand{\lc}{\left(}
\newcommand{\rc}{\right)}
\newcommand{\lv}{\left[}
\newcommand{\rv}{\right]}

\begin{document}

\title{{\it Ab initio} calculations with a nonspherical Gaussian basis set: 
Excited states of the hydrogen molecule}
\author{T.Detmer, P. Schmelcher and L. S. Cederbaum}
\address{Theoretische Chemie, Physikalisch--Chemisches Institut,\\
Universit\"at Heidelberg, INF 253, D--69120 Heidelberg,\\
Federal Republic of Germany}
\maketitle

\begin{abstract}
A basis set of generalized nonspherical Gaussian functions (GGTOs) is presented and discussed. 
As a first example we report on Born-Oppenheimer energies of the hydrogen molecule. 
Although accurate results have been obtained, we conclude that $H_2$ is too "simple" 
to allow for a substantial gain by using nonspherical functions. 
We rather expect that these functions may be particularly useful in calculations on large systems. 
A single basis set of GGTOs was used to simultaneously calculate the potential energy curves of 
several states within each subspace of $^{1,3}\Sigma_{g,u}$ symmetry. 
We hereby considerd the entire region of internuclear distances $0.8 \leq R \leq 1000\;a.u.$ 
In particular the results for the fourth up to sixth electronic states show a high accuracy compared to calculations 
which invoke explicitely correlated functions, e.g. the relative accuracy is at least 
of the order of magnitude of $10^{-5}\;a.u.$ 
Energies for the $4\,^1\Sigma_u^+$ and $4-6\,^3\Sigma_u^+$ were improved and 
accurate data for the $6\,^3\Sigma_g^+$, $5\,^1\Sigma_u^+$, and $6\,^1\Sigma_u^+$ state 
are, to the best of the authors knowledge, presented for the first time. 
Energy data for the seventh up to the nineth electronic state within each subspace were 
obtained with an estimated error of the order of magnitude of $10^{-4}\;a.u.$ 
The $7\,^1\Sigma_g^+$ and the $6\,^1\Sigma_u^+$ state were found to exhibit a very broad deep outer well 
at large internuclear distances.

\end{abstract}


\section{Introduction}
The usual method to obtain a molecular wavefunction is to expand 
this wavefunction in terms of products of linear combinations of atomic orbitals 
respecting the spatial and spin symmetries and the Pauli principle. 
In order to limit the computational expense a fast convergence of the calculation 
is desirable. The convergence  
essentially depends on the choice of the atomic basis set. 
Slater-type functions (STOs) and Gaussian-type functions (GTOs) hereby have 
gained large acceptance in quantum chemistry. 
Since computer evaluation of three- and four-center integrals over STO basis 
functions is very time consuming, Boys \cite{boys:1950} 
proposed in 1950 the use of GTOs instead of STOs. For GTOs 
the integral evaluation requires much less computer time than for STOs. 
This results from the fact that three- and four-center integrals over GTOs may 
be reduced to two-center-integrals. 

The slow convergence of the calculations is the major drawback of using GTOs. 
On the one hand, GTOs provide a poor description of the desired cusp 
of atomic orbitals for electrons located near the nucleus and the correlation cusp for small interelectronic distances. 
On the other hand, GTOs are not well suited for describing 
electronic states with distorted wavefunctions, e.g. for states involving high 
angular momenta. As a consequence, a linear combination of several GTOs is 
necessary for an accurate representation of such a wavefunction. 

A possible way of convergence improvement is to explicitely include the 
interelectronic distance in the wavefunction which was introduced by Hylleraas \cite{hylleraas} 
for the helium atom. For the hydrogen molecule the use of confocal elliptical coordinates 
was introduced by James and Coolidge \cite{james:1933} and proved to be very efficient. 
Further development on the latter method was done by Kolos and Wolniewicz (see e.g. Refs. \CITE{wolniewicz:1995,kolos:1994} 
for accurate ground state energies) which lead to 
theoretical energies for low lying electronic states 
of the hydrogen molecule with spectroscopic accuracy. 
However, the use of basis sets with confocal orbitals is restricted to diatomic molecules 
and cannot be used in polyatomic calculations. 

Recently, a basis set of generalized GTOs \cite{schmelcher3:1988} 
was introduced for calculations in the presence of external magnetic fields. 
This basis set of generalized (nonspherical) Gaussian functions (GGTOs) 
proved to be an excellent choice for calculations in strong magnetic fields. 
The flexibility of the basis functions allows for an adaption of the basis functions 
to the symmetry of the electronic wave function of the molecule which made the new GGTOs superior to 
simple GTOs or STOs. In the presence of an external magnetic field 
the wavefunction is, depending on the field strength, 
distorted towards a cylindrical symmetry. 
The obtained results for calculations on $H_2^+$ and $H_2$ in the presence of magnetic fields 
(see Refs. \CITE{kappes2:1996,detmer1:1998} and references therein) 
are very encouraging. This motivated 
the present authors to raise the question whether this basis set 
may be also useful for calculations in field free space. 
The first part of this paper deals with the presentation of the basis set 
and discusses briefly the possible advantages GGTOs may have in calculations on large systems. 
In the second part of this paper we report on a simple example: the investigation of 
the electronic structure of the hydrogen molecule using GGTOs.  
A detailed knowledge of this simplest diatomic molecule is of 
fundamental interest in different branches of chemistry and physics. 
Fundamental molecular processes can be studied, as , for example, 
charge exchange processes in $H^+ + H^-$ collisions , associative ionization reactions 
$H + H^* \rightarrow H + H^+ + e^-$, excitation reactions 
$H\lc1s\rc + H\lc1s\rc \rightarrow H\lc1s\rc + H\lc2s\rc$ or $H\lc1s\rc + H\lc2p\rc$ 
or chemical exchange processes $H + H_2 \rightarrow H_2 + H$. 
For a detailed understanding of the above mentioned processes one first needs 
accurate electronic energies for the hydrogen molecule. 
The investigations dealing with electronic states of $H_2$ , 
both experimentally and theoretically, are too numerous to be mentioned here. 
For an overview of experimental studies we refer the reader to 
Refs. \CITE{jungen:1990,rottke:1992} and references therein. 
In particular we mention an experiment recently reported by Reinhold et.al. \cite{reinhold:1997} 
where the outer minimum of the third excited  $4\,^1\Sigma_g^+$ state 
was investigated. 
For theoretical data on the electronic structure of $H_2$ see Refs. 
\CITE{kolos:1994,wolniewicz2:1993,wolniewicz:1998,dressler:1995,borondo:1987,liu:1994} 
and references therein. 

In the present investigation, a full CI approach is used to obtain the $n\,^{1,3}\Sigma_{g,u}^+\,n=1-9$ 
electronic states. Particular emphasize was put on accurate results for the first five excited states 
within each subspace of $\Sigma$ symmetry, e.g. the $n\,^{1,3}\Sigma_{g,u}^+\,n=2-6$ states. 
Our aim was to compute all of these states covering the entire range of internuclear distances 
from $R = 0.8\,a.u$ to $R = 1000\,a.u$  with a single basis set of GGTOs. 
In the present investigation we report on accurate Born-Oppenheimer energies of the 
third, fourth and  fifth excited $^1\Sigma_u^+$
and $^3\Sigma_u^+$ states and of the fifth excited $^3\Sigma_g^+$ state. 
Born-Oppenheimer energies for the third excited $^1\Sigma_u^+$ state have been given 
by Dressler and Wolniewicz \cite{dressler:1995}. 
These results were obtained with the aid of a basis set not optimized 
for the state in question and could be further improved. 
Excited $^3\Sigma_u^+$ states were investigated by Borondo et.al \cite{borondo:1987} 
but only with a very small basis set of GTOs and at a few selected internuclear distances. 
To the best of the authors knowledge Born-Oppenheimer energies of the  
$5 ^1\Sigma_u^+$, $6 ^1\Sigma_u^+$ and $6 ^1\Sigma_g^+$ states are presented here for the first time. 
The results on the other $n\,^{1,3}\Sigma_{g,u}^+\,n=1-6$ $H_2$ electronic states are compared 
with the best data found in the literature. 
As a side product of our calculations we obtained electronic energies for the $n\,^{1,3}\Sigma_{g,u}^+\,n=7-9$ states 
which have not been investigated up to now. These states are shown graphically but no energy data are provided.

\section{Generalized Gaussians}

The present computations were done with a basis set of generalized GTOs \cite{schmelcher3:1988} which has been 
originally introduced in order to perform calculation in the presence of an external magnetic field. 
In the most general appearance the basis functions in Cartesian coordinates read as follows:

\begin{eqnarray}
\phi_n\lc\bbox{r};\underline{\alpha},\bbox{R},\bbox{C}\rc &=& \exp\lv-i\bbox{A}\lc\bbox{C}\rc\bbox{r}\rv 
\lc x-R_x \rc ^{n_x} \lc y-R_y \rc ^{n_y} \lc z-R_z \rc ^{n_z}  \nonumber \\ && \times
\exp\lv-\lc\bbox{r}-\bbox{R} \rc^T \underline{\alpha} \lc\bbox{r}-\bbox{R} \rc\rv \label{aobas_a}
\end{eqnarray}
$\bbox{r} = \lc x,y,z \rc^T$ denotes the vector of the electronic ccordinates 
and $\bbox{R} = \lc R_x,R_y,R_z \rc^T$ is the position vector characterizing the center of the orbital. 
$\bbox{A}\lc \bbox{C} \rc$ ist the vector potential at the position $\bbox{C}$, where $\bbox{C}$ is 
a vector of variational parameters which are determined by minimizing the expectation value of the energy. 
The gauge factor $\exp\lv-i\bbox{A}\lc\bbox{C}\rc\bbox{r}\rv $ ensures the approximative gauge invariance of the 
energy expectation values. For our calculation on the $H_2$ molecule in the absence of a magnetic field 
this phase factor vanishes. 
For a detailed discussion of the dependence of energy expectation values on the choice of gauge 
we therefore refer the reader to Ref. \CITE{schmelcher3:1988}. 
Without the additional gauge factor needed in the presence of a magnetic field this type of basis functions 
was first introduced by Singer \cite{singer:1960}. 
The matrix 
\begin{equation}
\underline{\alpha} = \lv \begin{array}{ccc} 
\alpha_{xx} &  \alpha_{xy} & \alpha_{xz} \\
\alpha_{yx} &  \alpha_{yy} & \alpha_{yz} \\
\alpha_{zx} &  \alpha_{zy} & \alpha_{zz} \end{array} \rv
\end{equation}
is a real symmetric matrix of variational parameters which had to be optimized 
to minimize the energy expectation value. 

Basis functions of type (\ref{aobas_a}) possess several advantages compared to GTOs. 
In order to describe wave functions which differ from a pure spherical symmetry 
the number of basis functions may be reduced. 
Contrary to GTOs the flexibility of basis functions of type (\ref{aobas_a}) 
provide the possibility for an adaption of the basis 
functions according to the symmetry of the molecule being investigated. 
Therefore a more rapid convergence with increasing size of the basis set 
in energy calculations is expected  when dealing with 
wavefunctions which significantly differ from a spherical symmetry. 
The GGTO basis set can also be useful in calculations involving effects as angular correlation. 
If GTOs are used this leads, in particular for larger systems 
which involve high angular momenta, to a strong increase in the size of the basis set. 
In that situation the use of GGTOs might be very useful: 
As a consequence of the nonspherical shape, simple GGTOs contain portions of several different 
angular momenta and may therefore reduce the number of GTOs 
with high angular momenta significantly.  
Despite the obvious advantages of the GGTO basis compared to the GTO basis a main drawback is 
the evaluation of the matrix elements: The use of the GGTO basis requires a numerical integration 
in the evaluation of three- and four-center matrix elements. Hereby three-center matrix elements can 
be fastly evaluated but the evaluation of four-center matrix elements is time consuming. 
Consequently, for larger systems a combination of GTOs and GGTOs seems to be useful. 
For a description of core and other localized electrons GTOs are well suited 
since possible anisotropies in the wavefunction are small. 
GGTOs should be taken for a description of outer electrons and the electrons participating 
in the chemical bond and/or in those cases where angular correlation effects are relevant. 

As an example we considered the ground state of $H_2$ $1\,^1\Sigma_g^+$ with a total energy 
at the equilibrium distance of $-1.1744757\,a.u.$ \cite{wolniewicz:1995} and investigated the influence 
of the anisotropy of the basisfunction with the aid of  a (6s,3p,2d,1s) basis set. 
The energy was obtained by an optimization of the parameters for the GTO and GGTO basis  
at the CI level for $H_2$. For the GTO basis we obtained an energy of $-1.174006\;a.u.$ at the equilibrium distance. 
The use of anisotropic basis functions lead to an improved energy of $-1.174086\;a.u.$ 
The definition of the anisotropy $a$ is as follows: Let $\alpha_{xx} \leq \alpha_{zz}$, so $a := 1-\frac{\alpha_{zz}}{\alpha_{xx}}$. 
The anisotropy for s-type functions is pretty small, e.g., between $0.01$ and $0.05$. 
However it increased for functions involving angular momenta not equal zero $-$  
for these type of functions anisotropies up to $0.35$ were calculated. 
Compared to the GTO basis, the use of GGTOs yielded $17\,\%$ of the remaining energy to the exact result.

\section{Potential energy curves of $H_2$}
For an investigation of the electronic structure of the hydrogen molecule 
the basis functions in Eq. (\ref{aobas_a}) can be simplified significantly . 
Due to the cylindrical symmetry of the molecule we choose all off-diagonal elements 
of the matrix $\underline{\alpha}$ 
to be zero and in addition $\alpha_{xx}$ equal to $\alpha_{yy}$. 
In this work we used only orbitals centered at the positions of the nuclei. 
The origin of our coordinate system coincides with the midpoint of the internuclear axes 
and the protons are located on the z-axis. Therefore, 
the basis functions used in the present investigation read as follows: 
\begin{equation}
\phi_n \left(\bbox{r};\alpha_{xx},\alpha_{zz},\pm R/2 \right) = 
x^{n_x} y^{n_y} \left(z\mp R/2\right)^{n_z} \exp\left\{- \alpha_{xx}
\left(x^2 + y^2\right) - \alpha_{zz}\left( z\mp R/2 \right)^2 \right\} \label{aobas_s} ,
\end{equation}
For a detailed description of the evaluation of the various matrix elements 
needed in the computation of the wavefunction using this particular type of basis functions 
we refer the reader to Ref. \CITE{detmer:1997}. 

In the present study Born-Oppenheimer energies were calculated for the lowest nine states 
within each of the $\Sigma$ subspaces, e.g., for both singlet and triplet as well as gerade 
and ungerade parity. Hereby particular emphasize was put on accurate results for the $n\,^{1,3}\Sigma_{g,u}^+\,n=2-6$ states. 
A single basis set of GGTOs has been used for all states and internuclear distances. 
The determination of the electronic potential energy curves (PECs) was done by 
the following procedure:  Since we would like to perform calculations 
for a large range of internuclear distances, i.e. for $0.8\;a.u. < R < 1000\;a.u.$, we had 
to ensure obtaining correct energies in the dissoziation limit. Therefore, several GGTOs were optimized for 
electronic states of atomic hydrogen and included in the basis set.
Second, a limited number of basis functions for the description of angular correlation 
was optimized at a CI level at selected internuclear distances. 
Following this procedure we arrived at a number of approximately 3800 two particle functions at the full 
CI level of expansion. Most of these function were needed for the correct description of the 
dissociative behaviour. The anisotropies in the s-type functions varies drastically. 
S-type functions involving high parameter values (corresponding to electrons located near the nucleus)
show an almost negligible anisotropy  but functions with small parameter values 
(necessary for the description of higher excited states) possess anisotropies up to $1$. 
Basis functions describing angular correlation may possess even larger distortions. In our calculations 
the maximum of distortion was found to be $3$. 

In spite of the larger convergence error for energies of the $n\,^{1,3}\Sigma_{g,u}^+\,n=7-9$ states 
we graphically show the PECs for these states. 
The relative error in the dissociation limit varies between \tim4 and \tim6 
and is estimated to be of the order of magnitude of \tim4 in the vicinity of the 
equilibrium internuclear distance. Our results represent the first {\it ab initio } data for these excited states. 
The results allow us, in particular, to demonstrate the interaction of the 
attractive $H^+ + H^-$ states with $H(1s) + H(nl)$ states at very large internuclear distances. 
For this purpose a large regime of internuclear distances is necessary. 
The electronic energies were calculated at 480 different internuclear distances in the interval 
$0.8\;a.u. < R < 1000\;a.u.$ A complete table of the results can be 
obtained from the authors upon request.
The accuracy with respect to the energy data obtained in our investigation for states up to the fifth 
excited ones is estimated to be typically 
of the order of magnitude of \tim4 for $1\,^{1}\Sigma_{g,u}^+$ and \tim5 to \tim6 for the 
$n\,^{1}\Sigma_{g,u}^+\,n=2-6$ and $n\,^{3}\Sigma_{g,u}^+\,n=1-6$ states, respectively. 
The position of minima and maxima were determined 
with an accuracy of $10^{-2}\;a.u.$ of the internuclear distance. 

\subsection{The $^1\Sigma_g^+$ subspace}

Many calculations have been performed on Born-Oppenheimer states within 
the $^1\Sigma_g^+$ subspace. Very accurate energies for the ground state ($1\,^1\Sigma_g^+$) are given in Ref. \CITE{kolos:1994} 
and the lowest five excited states were considered in Ref. \CITE{wolniewicz:1993}. 
A very detailed theoretical investigation of the $4\,^1\Sigma_g^+$ was recently done by 
Wolniewicz \cite{wolniewicz:1998} in order to explain the experimental results 
by Reinhold et.al. \cite{reinhold:1997} The calculations mentioned above served as a benchmark 
in the determination of the accuray of our data. 
Our aim is the study of excited $H_2$ states and the data concerning the ground state 
of the hydrogen molecule is only a side product and less accurate than that for excited states. 
At the equilibrium internuclear distance of the ground state we obtained an energy of 
$1.1742937\;a.u.$ which implies an accuracy of $1.82 \ttim4 a.u.$ 
Our data for the $n\,^1\Sigma_g^+\,n=2-6$ states are in 
general at least as accurate as $5 \ttim5 a.u.$ or, equivalently, $\approx 11\,cm^{-1}$ 
compared to the results of Refs. \CITE{wolniewicz:1993,wolniewicz:1998}.  
For some internuclear distances our calculations for higher excited states, 
e.g. the $4\;^1\Sigma_g^+$ and $5\;^1\Sigma_g^+$ state, yield energies slightly lower than those 
given in Ref. \CITE{wolniewicz:1993}. The maximum difference in energy 
amounts to $4.229\;cm^{-1}$ at $R = 2.00\;a.u.$ ($E = -0.63490478\;a.u.$) for the $4\;^1\Sigma_g^+$ 
and $11.551\;cm^{-1}$ at $R = 2.80\;a.u.$ ($E = -0.61390242\;a.u.$) 
for the $5\;^1\Sigma_g^+$ state, respectively. 
These results indicate that a more accurate investigation of these excited states is still posible 
which can be done by an optimization of the wavefunction for these particular states. 

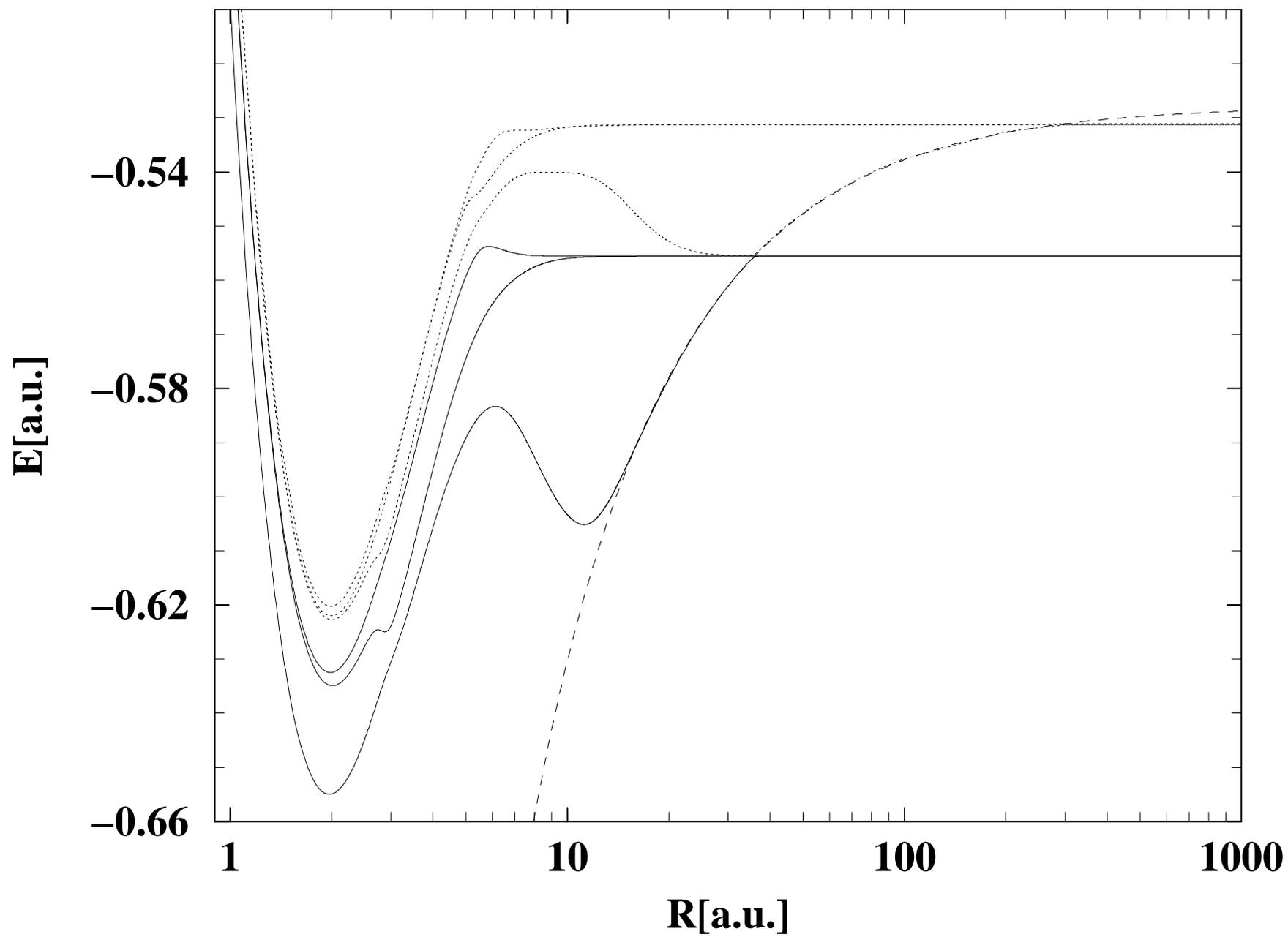
\begin{figure}
\caption{PECs (total energy in a.u.) for the third up to fifth (solid lines) and 
sixth up to eighth (dotted lines) excited $^1\Sigma_g^+$ state. To guide the eye we also included the $H^+ + H^-\lc 1s \rc^2$
diabatic state (dashed line).}
\label{sg_expap}
\end{figure}

The PECs for the $n\,^1\Sigma_g^+\,n=4-9$ states are shown in Fig. \ref{sg_expap}. 
In general we depict excited states up to the fifth one with solid and higher excited states 
with dotted lines. The figure nicely demontrates the series of avoided crossings 
between the corresponding $H(1s) + H(nl)$ Heitler-London configuration 
with the $H^+ + H^-\lc 1s \rc^2$ ion-pair configuration. 
The total ionic energy can be written \cite{wolniewicz:1998} as 
\begin{equation}
E_{ion}\lc R \rc = -0.527751014 - \frac{1}{R} -\frac{211.897}{\left(2R\right)^4}
\end{equation}
The corresponding PEC was also included in Fig. \ref{sg_expap}. For the $4\,^1\Sigma_g^+$ state its 
Born-Oppenheimer curve and the PEC of $H^-$ was found to be very close\cite{wolniewicz:1998}  
for internuclear distances $20 \leq R \leq 35.7$. 
The difference between higher excited states and the PEC of $H^-$ was not investigated up to now. 
The PEC of the $7\,^1\Sigma_g^+$ state 
is for some range of internuclear distances also very close to the PEC of $H^-$. 
In particular we mention that the $7\,^1\Sigma_g^+$ state exhibits a very broad deep outer well 
with a total energy at the minimum of $0.555487\;a.u.$ at large internuclear distances 
due to the interaction with the $H^+ + H^-\lc 1s \rc^2$ configuration. 
The depth of this well, e.g. the difference between the
maximum at $R = 8.7\;a.u.$ and the minimum at $R = 33.7\;a.u.$ approximately amounts to $0.015473\;a.u.$ 
A series of avoided crossings leads to the energetically equal dissociation limits of the 
$n\,^1\Sigma_g^+\,n=7-9$ states. The dissociation limit of the $10\,^1\Sigma_g^+$ state 
is the $H^+ + H^-\lc 1s \rc^2$ ion-pair configuration. 

\subsection{The $^1\Sigma_u^+$ subspace}

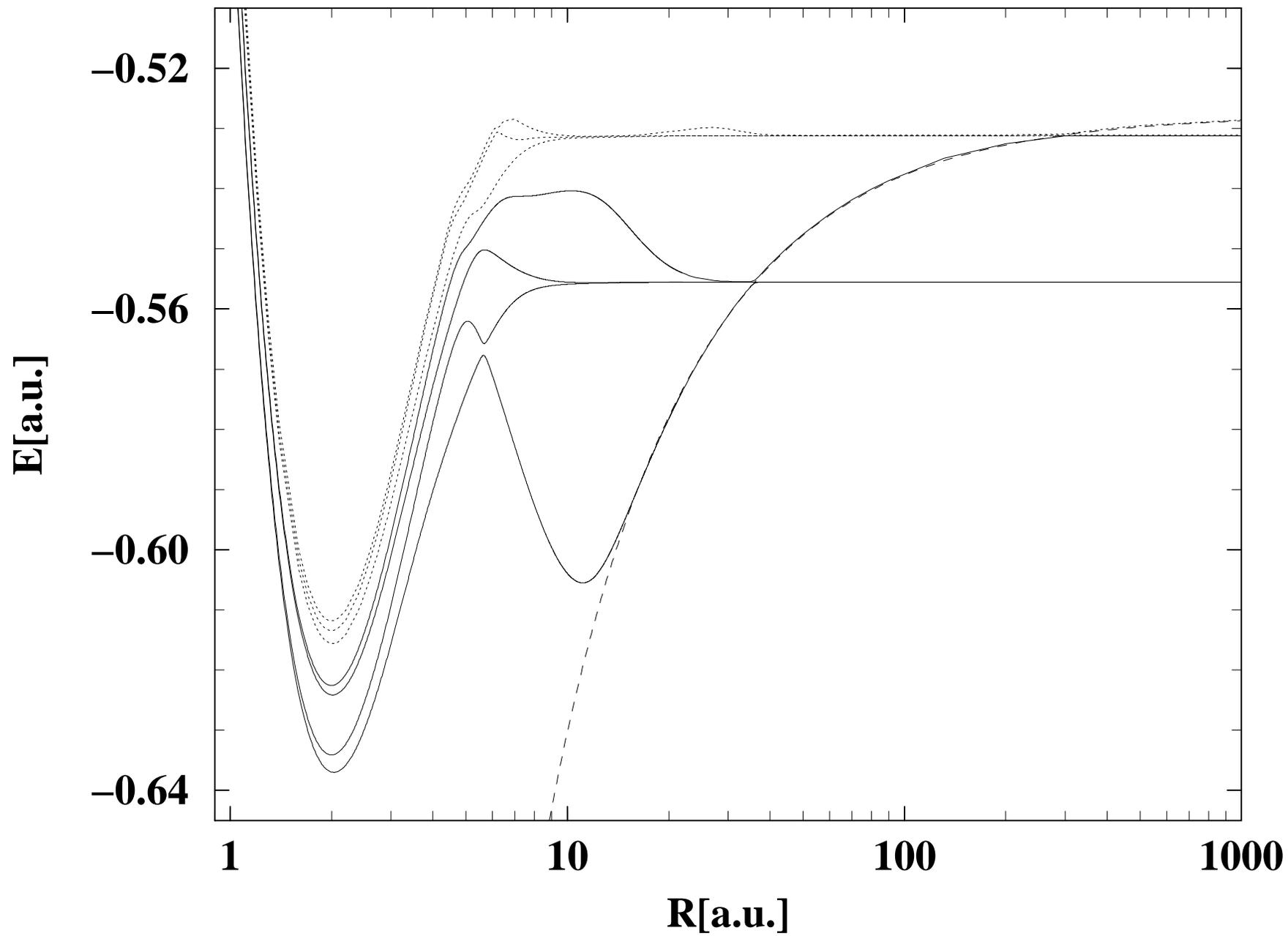
\begin{figure}
\caption{PECs (total energy in a.u.) for the second up to fifth (solid lines) and 
sixth up to eighth (dotted lines) excited $^1\Sigma_u^+$ state. To guide the eye we also included the $H^+ + H^-\lc 1s \rc^2$
diabatic state (dashed line).}  
\label{su_expap}
\end{figure}

For a reference of the most accurate data on the electronic energies of the 
four lowest states of   
$^1\Sigma_u^+$ symmetry we used the results given by Dressler and Wolniewicz \cite{dressler:1995}. 
In that investigation energies of the three lowest $^1\Sigma_u^+$ states were presented  
with high accuracy. Our results show a relative accuracy of the order of magnitude of 
\tim4 for the lowest and \tim5 for the first and second excited state. 
Energies for the $4\;^1\Sigma_u^+$ state however, were obtained in Ref. \CITE{dressler:1995} 
with the aid of a wavefunction optimized for the three lowest states. 
Therefore, these results are not optimal and can be improved. 
Table \ref{etab_ssu} lists our first time results on energy data for the fourth, fifth and sixth excited 
$^1\Sigma_u^+$ state at a few selected internuclear distances and a comparison 
of the present results with that of Ref. \CITE{dressler:1995} is given in Tab. \ref{comp_su}. 
From Tab. \ref{comp_su} we observe that a substantial improvement of the energy data of Ref. \CITE{dressler:1995} 
was possible at relevant internuclear distances. Hereby the most significant improvement of previous 
results $\lc 54.87\;cm^{-1}\rc $ was obtained in the vicinity of the 
(first) internuclear equilibrium distance at $R = 2.00\;a.u.$ 
The PECs of the $n\;^1\Sigma_u^+\,n=3-9$ states are graphically shown in Fig. \ref{su_expap}. 
Similarly to the $^1\Sigma_g^+$ subspace we recognize the existence of two deep outer wells 
at large internuclear distances. As in the case of the $^1\Sigma_g^+$ subspace this results from avoided 
crossings with the $H^+ + H^-\lc 1s \rc^2$ configuration. 
For the $3\;^1\Sigma_u^+$ state the existence of a second outer minimum 
was first predicted by Dabrowski and Herzberg \cite{dabrowski:1974} 
and theoretically shown by Kolos \cite{kolos:1976}. 
Our calculations revealed a broad and deep outer well of the $6\,^1\Sigma_u^+$ state  
which is located at $33.7\;a.u.$ and possesses a total energy of $-0.555492\;a.u.$ 
The depth of the well, e.g., the difference between the second minimum and the maximum of the 
$6\,^1\Sigma_u^+$ state amounts to $0.015134\;a.u.$

\subsection{The $^3\Sigma_g^+$ subspace}

Detailed investigations of triplet states can be found for instance in Refs. \CITE{kolos:1996,kolos:1990,bishop:1981}. 
Recently energy data for triplet states of $H_2$ for small internuclear distances were 
calculated by Liu and Hagstrom \cite{liu:1994} 
using full CI and large elliptical basis sets. 
The results given by Liu and Hagstrom \cite{liu:1994} and Kolos \cite{kolos:1996} 
served as the benchmark for our data for internuclear distances 
$R < 5\;a.u.$  and $R > 5\;a.u.$, respectively. 
First, we notice that our results for the $n\,^3\Sigma_{g,u}^+\;n=1-6$ triplet states 
(both gerade as well as ungerade symmetry) are roughly one order of magnitude more accurate 
than the corresponding singlet states, e.g. the relative accuracy is of the order of magnitude of \tim5 or \tim6. 
An explanation herefore is that 
for singlet states we deal with an electron cusp problem at $r_1 = r_2$ 
and therefore basis functions explicitely including the interelectronic distance should show a much faster 
convergence in particular at small internuclear distances. 
This cusp problem is absent for the triplet states. 
Furthermore, for triplet states correlation effects are reduced due to the Rydberg character of the states. 
As a consequence of our main goal - the accurate description of excited states - our basis set 
contains several s-type functions but 
only a few functions explicitely involving higher angular momenta. 
This further explains the better accuracy we have achieved for the triplet states.

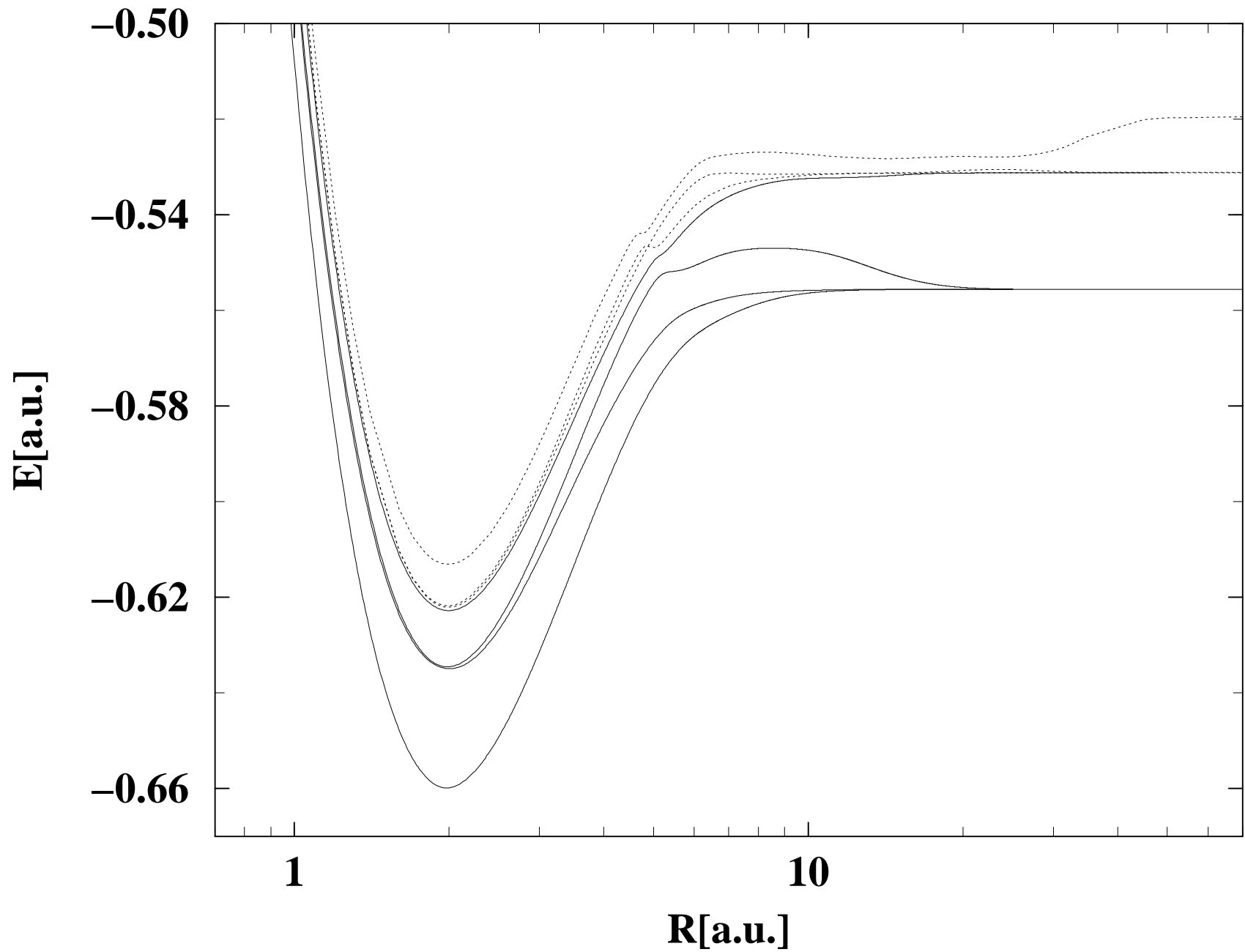
\begin{figure}
\caption{PECs (total energy in a.u.) for the second up to fifth (solid lines) and 
sixth up to eighth (dotted lines) excited $^3\Sigma_g^+$ state.}  
\label{tg_expap}
\end{figure}

For the four lowest $^3\Sigma_g^+$ states our results show an accuracy of the order of magnitude of \tim6 
in the vicinity of the equilibrium internuclear distance and in the dissociation limit and of \tim5 
for intermediate internuclear distances, e.g. $5 \leq R \leq 10\;a.u.$ 
In this work we report for the first time results of accurate energies for the $6\,^3\Sigma_g^+$ state. 
The corresponding data are given in Tab. \ref{etab_stg} and the PECs are shown in Fig. \ref{tg_expap}. 
For the $(5\,^3\Sigma_g^+)$ state a comparison of our results with 
that of Refs. \CITE{kolos:1996,liu:1994} is presented in Tab. \ref{comp_tg}. 
For internuclear distances near the equilibrium configuration our energy data are 
slightly above that of Liu and Hagstrom but for larger distances slightly better than 
the best previous results given in Ref. \CITE{kolos:1996}. 
Figure \ref{tg_expap} indicates that the $(7\,^3\Sigma_g^+)$ and $(8\,^3\Sigma_g^+)$ states 
(similiar to the $(4\,^3\Sigma_g^+)$ and $(5\,^3\Sigma_g^+)$ states) 
are nearly degenerate near the equilibrium internuclear distance. 
However, the energies for higher excited states ($n=7-9$) are not as accurate as those for the 
$(4\,^3\Sigma_g^+)$ and $(5\,^3\Sigma_g^+)$ states and therefore a more detailed 
study has to be performed in order to confirm this result. 

\subsection{The $^3\Sigma_u^+$ subspace}

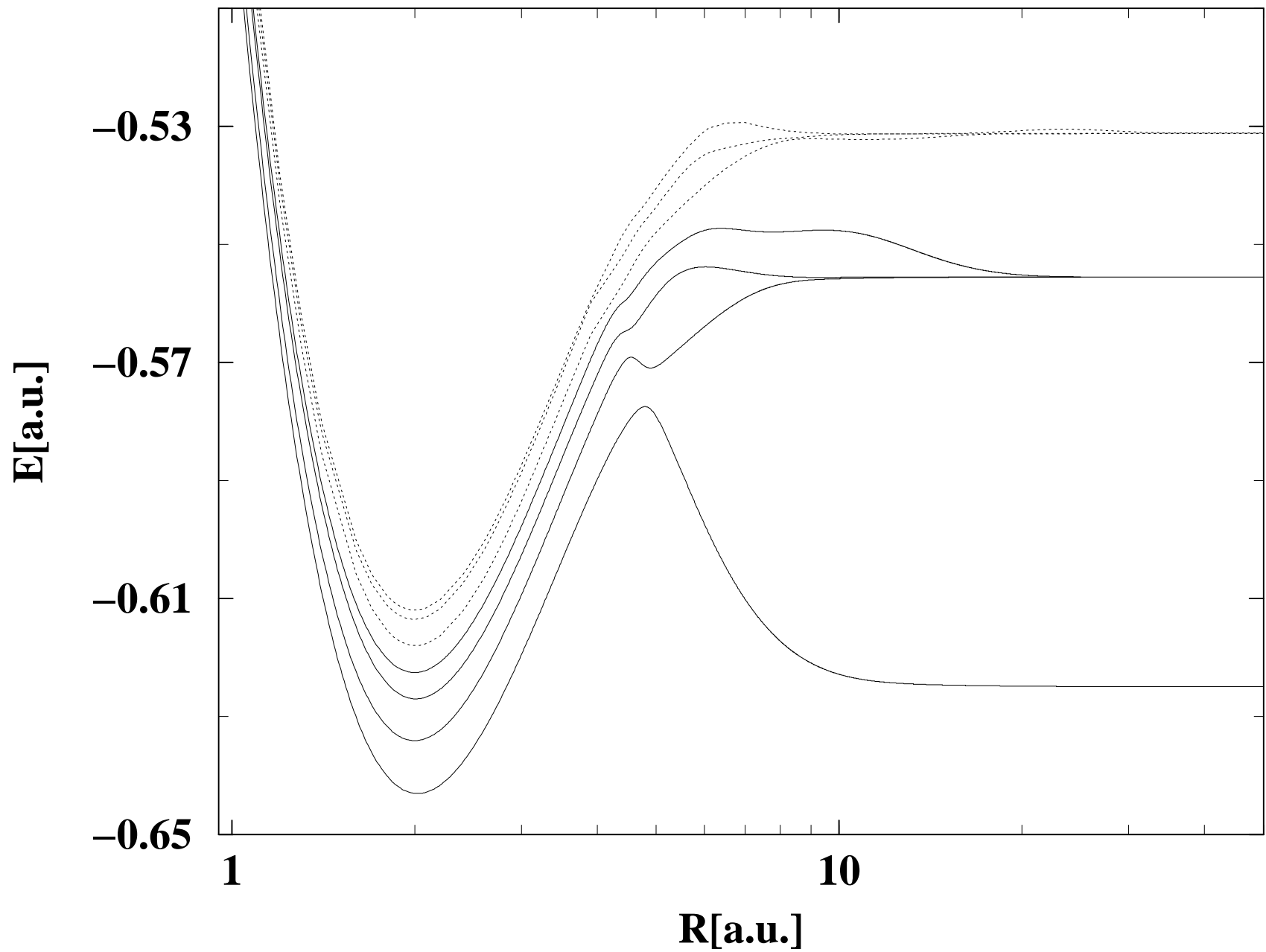
\begin{figure}
\caption{PECs (total energy in a.u.) for the second  up to fifth (solid lines) and 
sixth up to eighth (dotted lines) excited $^3\Sigma_u^+$ state.}  
\label{tu_expap}
\end{figure}

For accurate energy data concerning the three lowest states of the $^3\Sigma_u^+$ subspace 
we refer the reader to Refs. \CITE{liu:1994,kolos:1994} and references therein. 
A qualitative description of the six lowest $^3\Sigma_u^+$ states can also be found in the 
work by Borondo et.al. \cite{borondo:1987} 
However, in that paper energy data for the six lowest $^3\Sigma_u^+$ states 
were obtained using a small GTO basis and the energy data is therefore amenable to drastic improvement. 
Our results for the three lowest $^3\Sigma_u^+$ states show a typical accuracy of \tim6 compared to 
the best data given in Refs. \CITE{liu:1994,kolos:1994}. 
For the $3\;^3\Sigma_u^+$ state and some particular internuclear distances 
the present calculations yield energies which are 
several tenths of $cm^{-1}$  lower than those reported in Ref. \CITE{kolos:1994}. 
Table \ref{etab_stu} lists our energy data for the $n\,^3\Sigma_{u}^+\;n=4-6$ 
states.  PECs are depicted in Fig. \ref{tu_expap}. In general our results are of the order 
of magnitude of \tim3 or \tim2 {\it lower} than that given by Borondo \cite{borondo:1987}. 
Again, we draw the readers attention to the series of avoided crossings originating from 
an interaction with an $H^+ + H^-\lc 1s2s\rc$ ion pair configuration \cite{borondo:1987}.  

\section{Conclusions}

In the first part of the paper we presented a basis set of generalized Gaussian type functions. This basis set 
is well suited for a description of distorted wave functions and angular correlation effects.  
$H_2$ has been investigated as a first test. For $H_2$ the basis set works well, but may not justify the larger effort 
involved in the use of nonspherical functions. We rather expect this type of functions to be useful in larger systems. 

We investigated the $n\,^{1,3}\Sigma_{g,u}^+\,n=1-9$ electronic states of the hydrogen molecule using 
one single basis set of GGTOs. 
A broad range of internuclear distances ($0.8 < R < 1000\;a.u.$) has been considered. 
Our main goal was the calculation of accurate Born-Oppenheimer energies for the $n\,^{1,3}\Sigma_{g,u}^+\,n=4-6$ 
states. For the first time we present PECs for the $n\,^{1,3}\Sigma_{g,u}^+\,n=7-9$ states. 
Energy data for the $4\,^1\Sigma_u^+$ and the $n\,^3\Sigma_u^+\,n=4-6$ were improved compared to the best data 
available in the literature. 
New accurate PECs were calculated for the $5 ^1\Sigma_u^+$, $6 ^1\Sigma_u^+$ and $6 ^1\Sigma_g^+$ state. 
The $7\,^1\Sigma_g^+$ and the $6\,^1\Sigma_u^+$ state were found to provide a second deep outer well 
arising due to an interaction with the diabatic $H^+ + H^-\lc 1s \rc^2$ state. 
Similarly to the $4\,^1\Sigma_g^+$ these states also may contain several long lived valence states of $H_2$ 
in analogy to what has been discussed in Ref. \CITE{wolniewicz:1998}.

\bibliographystyle{prsty}
\bibliography{magnet}

\begin{table}
\caption{Born-Oppenheimer energies E of the 4 $^1\Sigma_u^+$, 5 $^1\Sigma_u^+$ and 6 $^1\Sigma_u^+$ states. 
Dissociation energies D are in $cm^{-1}$ (1 a.u.= 219474.64 $cm^{-1}$), 
all other quantities are in atomic units}
{\squeezetable
\begin{tabular}{ddddddd}
\multicolumn{1}{c}{R} &
\multicolumn{1}{r}{$E\lc 4\,^1\Sigma_u^+\rc $} &
\multicolumn{1}{c}{$D\lc 4\,^1\Sigma_u^+\rc $} &
\multicolumn{1}{r}{$E\lc 5\,^1\Sigma_u^+\rc $} &
\multicolumn{1}{c}{$D\lc 5\,^1\Sigma_u^+\rc $} &
\multicolumn{1}{r}{$E\lc 6\,^1\Sigma_u^+\rc $} &
\multicolumn{1}{c}{$D\lc 6\,^1\Sigma_u^+\rc $}  \\\hline
   0.8 & -0.33576673 &  -48238.074 & -0.32449461 &  -50712.018 & -0.32427855 &  -45424.985  \\
   1.2 & -0.56030161 &    1041.638 & -0.54932463 &   -1367.529 & -0.54881235 &    3854.491  \\
   1.6 & -0.62232286 &   14653.730 & -0.61179855 &   12343.912 & -0.61081490 &   17462.477  \\
   1.8 & -0.63167582 &   16706.467 & -0.62142478 &   14456.625 & -0.62015441 &   19512.263  \\
   1.9 & -0.63354815 &   17117.397 & -0.62344232 &   14899.424 & -0.62201911 &   19921.517  \\
   2.0 & -0.63409815 &   17238.107 & -0.62414191 &   15052.966 & -0.62256086 &   20040.418  \\
   2.1 & -0.63362191 &   17133.585 & -0.62381875 &   14982.039 & -0.62207577 &   19933.953  \\
   2.2 & -0.63235117 &   16854.690 & -0.62270381 &   14737.339 & -0.62079556 &   19652.979  \\
   2.6 & -0.62246079 &   14684.003 & -0.61345778 &   12708.070 & -0.61086073 &   17472.537  \\
   3.0 & -0.60934757 &   11805.983 & -0.60104221 &    9983.167 & -0.59768853 &   14581.572  \\
   4.0 & -0.57870623 &    5080.986 & -0.57269612 &    3761.919 & -0.56675253 &    7791.905  \\
   5.0 & -0.56213226 &    1443.420 & -0.55458442 &    -213.140 & -0.54983071 &    4077.994  \\
   5.6 & -0.56549852 &    2182.229 & -0.55022956 &   -1168.921 & -0.54572663 &    3177.253  \\
   5.7 & -0.56571034 &    2228.718 & -0.55017260 &   -1181.422 & -0.54507868 &    3035.045  \\
   5.8 & -0.56517395 &    2110.994 & -0.55025441 &   -1163.466 & -0.54447672 &    2902.929  \\
   6.0 & -0.56372910 &    1793.886 & -0.55066494 &   -1073.366 & -0.54341151 &    2669.143  \\
   7.0 & -0.55863828 &     676.580 & -0.55321945 &    -512.716 & -0.54129338 &    2204.267  \\
   8.0 & -0.55673204 &     258.208 & -0.55465619 &    -197.387 & -0.54114524 &    2171.753  \\
  10.0 & -0.55589629 &      74.784 & -0.55554014 &      -3.383 & -0.54037628 &    2002.988  \\
  15.0 & -0.55561994 &      14.130 & -0.55558225 &       5.859 & -0.54599714 &    3236.623  \\
  20.0 & -0.55557306 &       3.841 & -0.55556074 &       1.138 & -0.55281366 &    4732.676  \\
  30.0 &             &             &             &             & -0.55542802 &    5306.462  \\
  35.0 &             &             &             &             & -0.55547294 &    5316.322  \\
  40.0 &             &             &             &             & -0.55260171 &    4686.158  \\
 100.0 &             &             &             &             & -0.53756117 &    1385.142  \\
 200.0 &             &             &             &             & -0.53256019 &     287.553  \\
 \multicolumn{1}{l}{\hspace*{2mm}$\infty$} & -0.55555555 &       & -0.55555555 &       & -0.53125000 &        \\
\end{tabular}
}
\label{etab_ssu}
\end{table}

\begin{table}

\caption{Comparison of Born-Oppenheimer energies for the $4\;^3\Sigma_u^+$ state 
with data given by Dressler \cite{dressler:1995}. 
Energy differences $\Delta$ = E(Lit.) - E(present) are in are in $cm^{-1}$, 
all other quantities are in atomic units}
\begin{tabular}{dddd}
\multicolumn{1}{c}{R} &
\multicolumn{1}{c}{E(Ref. \CITE{dressler:1995})} &
\multicolumn{1}{c}{E(present work)} &
\multicolumn{1}{c}{$\Delta$} \\\hline
  1.0  &  -0.482967317 & -0.48309139  & +27.20  \\ 
  2.0  &  -0.633848144 & -0.63409815  & +54.87  \\ 
  5.1  &  -0.561982447 & -0.56205148  & +15.15  \\ 
  5.7  &  -0.565729995 & -0.56571034  & -4.31   \\ 
 11.2  &  -0.555749330 & -0.55576957  & +4.44   \\ 
 20.0  &  -0.555531578 & -0.55557305  & +9.10   \\ 
 30.0  &  -0.555505107 & -0.55555535  & +11.03  \\ 
\end{tabular}
\label{comp_su}
\end{table}

\begin{table}
\caption{Born-Oppenheimer energies E of the 6 $^3\Sigma_g^+$ states. 
Dissociation energies D are in $cm^{-1}$ , 
all other quantities are in atomic units}

\begin{tabular}{ddd}
\multicolumn{1}{c}{R} &
\multicolumn{1}{r}{$E\lc6\,^3\Sigma_g^+\rc$} &
\multicolumn{1}{c}{$D\lc6\,^3\Sigma_g^+\rc$}  \\ \hline
   0.8  & -0.32578808 &  -45093.681    \\ 
   1.2  & -0.54971049 &    4051.610    \\ 
   1.6  & -0.61124180 &   17556.172    \\ 
   1.8  & -0.62043117 &   19573.006    \\ 
   1.9  & -0.62225080 &   19972.368    \\ 
   2.0  & -0.62276944 &   20086.196    \\ 
   2.1  & -0.62228292 &   19979.418    \\ 
   2.2  & -0.62102141 &   19702.547    \\ 
   2.6  & -0.61131929 &   17573.178    \\ 
   3.0  & -0.59856552 &   14774.050    \\          
   4.0  & -0.56903655 &    8293.189    \\        
   5.0  & -0.54974877 &    4060.012    \\ 
   6.0  & -0.54122673 &    2189.639    \\ 
   7.0  & -0.53619376 &    1085.031    \\ 
   8.0  & -0.53388710 &     578.776    \\ 
  10.0  & -0.53241705 &     256.138    \\ 
  12.5  & -0.53214863 &     197.226    \\ 
  15.0  & -0.53171747 &     102.597    \\ 
  20.0  & -0.53128098 &       6.800    \\ 
  30.0  & -0.53125008 &       0.018    \\
\multicolumn{1}{c}{$\infty$} & -0.53125000 &      0 \\
 \end{tabular}
\label{etab_stg}
\end{table}

\begin{table}

\caption{Comparison of Born-Oppenheimer energies for the 5 $^3\Sigma_g^+$ state 
with data given by Liu and Hagstrom \cite{liu:1994} ($R \leq 3.0\;a.u.$) and Kolos \cite{kolos:1996} ($R \geq 6.0\;a.u.$) 
Energy differences $\Delta$ = E(Lit.) - E(present) are in are in $cm^{-1}$, 
all other quantities are in atomic units}
\begin{tabular}{dddd}
\multicolumn{1}{c}{R} &
\multicolumn{1}{c}{E(Lit.)} &
\multicolumn{1}{c}{E(present work)} &
\multicolumn{1}{c}{$\Delta$} \\ \hline
  1.0  &  -0.483282334 & -0.48328606   &  +0.82   \\ 
  2.0  &  -0.634509581 & -0.63450920   &  -0.08   \\ 
  3.0  &  -0.607932413 & -0.60792794   &  -0.98   \\ 
  6.0  &  -0.550672651 & -0.550678881  &  +1.36   \\ 
 12.0  &  -0.549376938 & -0.549382559  &  +1.23   \\ 
 20.0  &  -0.555171208 & -0.555206182  &  +7.67   \\ 
\end{tabular}
\label{comp_tg}
\end{table}

\begin{table}

\caption{Born-Oppenheimer energies E of the 4 $^3\Sigma_u^+$, 5 $^3\Sigma_u^+$ and 6 $^3\Sigma_u^+$ states. 
Dissociation energies D are in $cm^{-1}$, 
all other quantities are in atomic units}
{\squeezetable
\begin{tabular}{ddddddd}
\multicolumn{1}{c}{R} &
\multicolumn{1}{r}{$E\lc 4\,^3\Sigma_u^+\rc $} &
\multicolumn{1}{c}{$D\lc 4\,^3\Sigma_u^+\rc $} &
\multicolumn{1}{r}{$E\lc 5\,^3\Sigma_u^+\rc $} &
\multicolumn{1}{c}{$D\lc 5\,^3\Sigma_u^+\rc $} &
\multicolumn{1}{r}{$E\lc 6\,^3\Sigma_u^+\rc $} &
\multicolumn{1}{c}{$D\lc 6\,^3\Sigma_u^+\rc $} \\\hline
   0.8 & -0.33576681 &  -48238.056 & -0.32612114 &  -50355.035 & -0.32430141 &  -50754.421  \\
   1.2 & -0.56030175 &    1041.670 & -0.55178019 &    -828.598 & -0.54882094 &   -1478.077  \\
   1.6 & -0.62232310 &   14653.783 & -0.61471484 &   12983.963 & -0.61081948 &   12129.030  \\
   1.8 & -0.63167611 &   16706.532 & -0.62437360 &   15103.815 & -0.62015808 &   14178.616  \\
   1.9 & -0.63354848 &   17117.468 & -0.62636467 &   15540.805 & -0.62202245 &   14587.798  \\
   2.0 & -0.63409850 &   17238.185 & -0.62701366 &   15683.242 & -0.62256394 &   14706.640  \\
   2.1 & -0.63362229 &   17133.668 & -0.62661900 &   15596.623 & -0.62207862 &   14600.126  \\
   2.2 & -0.63235157 &   16854.778 & -0.62541462 &   15332.294 & -0.62079822 &   14319.110  \\
   2.6 & -0.62246125 &   14684.104 & -0.61567539 &   13194.780 & -0.61086284 &   12138.546  \\
   3.0 & -0.60934867 &   11806.224 & -0.60259825 &   10324.678 & -0.59769020 &    9247.486  \\
   4.0 & -0.57869562 &    5078.657 & -0.57201487 &    3612.402 & -0.56686141 &    2481.348  \\
   4.8 & -0.56345828 &    1734.448 & -0.55734267 &     392.226 & -0.55131413 &    4403.567  \\
   4.9 & -0.57096403 &    3381.769 & -0.55979427 &     930.290 & -0.55402538 &     335.835  \\
   5.0 & -0.57068654 &    3320.867 & -0.55859069 &     666.136 & -0.55306333 &    -546.981  \\
   5.5 & -0.56730268 &    2578.195 & -0.55473890 &    -179.236 & -0.54957213 &   -1313.210  \\
   6.0 & -0.56397471 &    1847.791 & -0.55382444 &    -379.937 & -0.54764726 &   -1735.670  \\
   7.0 & -0.55910158 &     778.262 & -0.55465050 &    -198.636 & -0.54763133 &   -1739.167  \\
   8.0 & -0.55678031 &     268.802 & -0.55537691 &     -39.209 & -0.54789839 &   -1680.555  \\
  10.0 & -0.55583534 &      61.406 & -0.55559346 &       8.319 & -0.54769291 &   -1725.651  \\
  15.0 & -0.55559282 &       8.179 & -0.55553531 &      -4.444 & -0.55285806 &    -592.032  \\
  20.0 & -0.55556118 &       1.234 & -0.55554193 &      -2.989 & -0.55519059 &     -80.101  \\
  25.0 & -0.55555625 &       0.152 & -0.55554759 &      -1.748 & -0.55552310 &      -7.123  \\
\multicolumn{1}{c}{$\infty$} & -0.55555555 &      & -0.55555555 &  & -0.55555555 & \\
\end{tabular}
}
\label{etab_stu}
\end{table}

\end{document}